\documentclass[useAMS,usenatbib,usegraphicx]{mn2e}

\def\h2o{H$_{\rm 2}$O}
\def\ch4{CH$_{\rm 4}$}

\def\ca1{Ca\textsc{i}}
\def\na1{Na\textsc{i}}
\def\k1{K\textsc{i}}
\def\rb1{Rb\textsc{i}}
\def\ti1{Ti\textsc{i}}
\def\mn1{Mn\textsc{i}}
\def\al1{Al\textsc{i}}

\def\chisq{$\chi^2$}
\def\micron{$\mu$m}
\def\teff{${\rm T}_{\rm eff}$}
\def\cor{$\rho_{i,i'}$}
\def\modcor{$|\rho_{i,i'}|$}

\title{Correlated spectral variability in brown dwarfs}

\author[C.A.L.\ Bailer-Jones]
{C.A.L.\ Bailer-Jones\thanks{Email: calj@mpia-hd.mpg.de}\\ Max-Planck-Institut f\"ur Astronomie, K\"onigstuhl 17, D-69117
Heidelberg, Germany}

\begin{document}

\date{Submitted 2 March 2007; Accepted 27 November 2007}

\maketitle

\label{firstpage}

\begin{abstract}
  Models of brown dwarf atmospheres suggest they exhibit complex physical
  behaviour.  Observations have shown that they are indeed dynamic, displaying
  small photometric variations over timescales of hours. Here I report results
  of infrared (0.95--1.64\,\micron) spectrophotometric monitoring of four
  field L and T dwarfs spanning timescales of 0.1--5.5\,hrs, the goal being to
  learn more about the physical nature of this variability.  Spectra are
  analysed differentially with respect to a simultaneously observed reference
  source in order to remove Earth-atmospheric variations. The variability
  amplitude detected is typically 2--10\%, depending on the source and
  wavelength.  I analyse the data for correlated variations between spectral
  indices.  This approach is more robust than single band or \chisq\ analyses,
  because it does not assume an amplitude for the (often uncertain) noise
  level (although the significance test still assumes a shape for the noise
  power spectrum).  Three of the four targets show significant evidence for
  correlated variability. Some of this can be associated with specific
  features including Fe, FeH, VO and \k1, and there is good evidence for
  intrinsic variability in \h2o\ and possibly also \ch4.  Yet some of this
  variability covers a broader spectral range which would be consistent with
  dust opacity variations.  The underlying common cause is plausibly localized
  temperature or composition fluctuations caused by convection. Looking at the
  high signal-to-noise ratio stacked spectra we see many previously identified
  spectral features of L and T dwarfs, such as \k1, \na1, FeH, \h2o\ and \ch4.
  In particular we may have detected methane absorption at 1.3--1.4\,\micron\ 
  in the L5 dwarf SDSS\,0539$-$0059.

\end{abstract}


\section{Introduction}\label{introduction}

Numerous observational studies over recent years have revealed photometric
variability in old brown dwarfs and very low mass stars with spectral
types late M, L and T (collectively ultracool dwarfs, or UCDs) (e.g.\ 
Gelino et al.\ \cite{gelino02}, Koen et
al.\ \cite{koen04}). The variability is always of low amplitude (a few tens of
millimagnitudes) and often nonperiodic, leading some authors to interpret this
as a result of intrinsic atmospheric variability (Bailer-Jones \& Mundt
\cite{bjm01}) rather than a rotational modulation. The idea here is that,
while UCDs are rapid rotators (Mohanty \& Basri
\cite{mohanty03}, Bailer-Jones \cite{bj04}) with likely periods of 3--10\,hrs,
heterogeneous variations in opacity on a shorter timescale could account for
the observed signature.  In an earlier review of the literature, Bailer-Jones
\cite{bj05} estimated that 40\% of UCDs are variable, although this may be a
lower limit due to limited observations, plus the fact that a given UCD is not
always variable. For a more recent review and discussion see Goldman
\cite{goldman05}.

Among young (1--10\,Myr) low mass stars and brown dwarfs (with mid and late M
spectral types) variability is also common (e.g.\ Joergens et
al.~\cite{joergens03}), with also about $\sim 40$\% of objects variable, but
this is typically periodic and with larger amplitudes of up to a few tenths of
a magnitude (reviewed in Bailer-Jones \cite{bj05}). This is similar to that
seen in (low mass) T Tauri stars, where the variability is related to
cool spots.  For the older, cooler objects considered here the physical
mechanism behind the variability is less clear, because the predicted
neutrality of their atmospheres inhibits formation of magnetically-cooled
regions.  Local opacity variations could arise from local temperature or
composition variations, one or both of which could be caused from convective
flows of matter to/from the deeper, opaque regions of the atmosphere.

To investigate this further, multiband monitoring is required. Bailer-Jones
\cite{bj02} (hereafter BJ02) used synthetic UCD spectra to predict the
variability signature due to spots and dust clouds and used these to
analyse spectral time series of an L1.5 dwarf. While there was no evidence
for variability in a single channel, there was a suggestion of correlated
variations which were consistent with dust. Nakajima et al.\ \cite{nakajima00} found a
suggestion of water variability in a T dwarf, while Clarke et al.
\cite{clarke03} found no variability in the dust-sensitive band heads of TiO,
CrH and FeH of the L2V Kelu-1.  

Here I report on infrared spectrophotometric monitoring of four
UCDs.  This is an extension of a pilot study in BJ02. 
The objective is to identify which (if any) parts of the spectrum are most
variable and, moreover, which regions show significant correlated variability
with one another. If the correlated variations can be traced to specific
molecules (e.g.\ water, methane, FeH) or dust-sensitive regions, then not only
is this a more robust detection of intrinsic variability (as opposed to
instrumental or telluric artefacts), but may provide some clues to the nature
of the atmospheric variations.

\section{Data acquisition and reduction}\label{datared}

\subsection{Targets}

Targets were chosen based on their observability (RA and Dec), brightness and
spectral type.  There had to be a bright, nearby star to act as a reference
(but not too bright or close), although this was met by most candidates. The
targets are listed in
Table~\ref{targets}. 2MASS\,0559 was monitored by Enoch et al.\ \cite{enoch03} who
found no Ks-band variability above an amplitude of 0.1\,mag over a timescale
of a few days. Tinney et al.\ \cite{tinney03} measured its parallax and
suspected it to be a binary on account of its apparent overluminosity (i.e.\ 
compared to models), although Burgasser et al.\ \cite{burgasser03} found no
evidence for binarity from HST observations.  SDSS\,0539 was monitored in the
I-band by Bailer-Jones \& Mundt~\cite{bjm01} who found it to be significantly
variable (p\,=\,3e-5) at a period of $13.3\pm1.2$\,hrs (20$\sigma$ detection
in the periodogram).

\begin{table*}
\begin{minipage}{122mm}
\caption[]{Targets observed. The reference is the discovery paper and
provides the spectral type and full name.  The photometry is from 2MASS
from the archive compiled by Kirkpatrick \cite{kirkpatrick03}.
\label{targets}
}
\begin{tabular}{lllrrl}
\hline
name            & Full name                      & SpT   & J     & reference     \\
\hline
SDSS\,0539  & SDSSp J053951.99-005902.0  &  L5  &  14.03   & Fan et al.\  \cite{fan00}      \\
2MASS\,0559  & 2MASS J05591914-1404488    &  T5  &  13.80   & Burgasser et al.\ \cite{burgasser00} \\
SSSPM\,0828$^*$  & SSSPM J0829-1309           &  L2  &  12.80   & Scholz \& Muesinger \cite{scholz2002}\\
2MASS\,0847  & 2MASS J08472872-1532372    &  L2  &  13.51   & Cruz et al.\ \cite{cruz03} \\
\hline
\end{tabular}
$^*$~The discoverers (unconventionally) round off the target's RA coordinate
(which they report as J2000 08h\,28m\,34.11s) to the nearest minute in
assigning a name, rather than truncating it. I avoid this, resulting in the
shortened name SSSPM\,0828.
\end{minipage}
\end{table*}

\subsection{Observations}\label{observations}

Infrared spectra were obtained with SOFI on the 3.5m NTT ESO telescope on La
Silla, Chile. The field-of-view was 4.98$'$\,$\times$\,4.98$'$ and the pixel
scale 0.292$''$/pix. I used the blue grism with a dispersion of 0.696\,nm/pix
which imaged the spectral range 0.95--1.64\,\micron\ in first order.  With a
spectrograph slit width of 2$''$ this delivered a nominal resolving power of
300, corresponding to a spectral profile with full width at half maximum
(FWHM) of 3.3\,nm at 1\,\micron\ (5 pixels) and 5.3\,nm at 1.6\,\micron\ (8
pixels).

The observing procedure is similar to that used in BJ02.  Each target was
monitored continuously for a few hours, dithered between two positions on the
slit (separated by about 1 arcmin).  At each position a set of three 60s
exposure images was obtained (each comprising three co-added 20s exposures).
The total time required for the two consecutive dither positions (which will
form a ``paired spectrum'') is 7.8 minutes.  A nearby reference star of
comparable brightness was placed in the slit and spectra acquired
simultaneously.
Data were acquired over two nights in January 2004 as listed in
Table~\ref{obslog}.  Although mostly clear, some thin cloud was present at times.

\begin{table}
\caption[]{Observing log. The normal spacing between
  consecutive paired spectra is 7.8 minutes. The FWHM refers to the width of the
  extracted spectra (which is generally larger than the image seeing).
\label{obslog}
}
\begin{tabular}{llcll}
\hline
UT range       & name  & \# paired  &  airmass  & FWHM \\
               &       & spectra    &           & arcsec \\
\hline
\multicolumn{5}{c}{{\em Night 1 (2004 January 10 UT)}} \\
01:28 -- 03:42 & SDSS\,0539  & 16  & 1.14--1.20  & 0.9--1.3 \\ 
03:56 -- 09:22 & SSSPM\,0828  & 30  & 1.04--1.56  & 0.9 \\ 
\hline
\multicolumn{5}{c}{{\em Night 2 (2004 January 11 UT)}} \\
01:20 -- 05:06 & 2MASS\,0559  & 26 & 1.04--1.18  & 0.9 \\ 
05:30 -- 09:20 & 2MASS\,0847  & 25 & 1.04--1.39  & 0.9--1.5 \\  
\hline
\end{tabular}
\vspace*{1ex}
\end{table}

\subsection{Image reductions and spectral extraction}\label{imagered}

The three consecutive images at each of the two positions were summed. The
difference between these removes most of the sky background and gives the two
{\em difference images}, each with an effective exposure time of three
minutes.
After flat fielding (see below), the spectra were extracted 
with the {\sc APSUM} package in IRAF. A residual background
on either side of the spectrum is computed and subtracted.  The spectra are
wavelength calibrated using Xenon arc lamps.
A 16-line fit with a cubic
Chebyshev polynomial produced an RMS fit of 0.13\,nm, which is 
about 1/30 the spectral FWHM.  
Each spectrum is cubic spline interpolated to a uniform wavelength scale
(0.696\,nm/pix) with 1028 pixels.  The spectra are not flux calibrated and the
flux scale is proportional to photon counts.

Because the seeing disk was typically smaller than the slit width, successive
spectra could show small shifts in the dispersion direction.
This would produce variations in the wavelength zero point.
These I determined by cross correlating each extracted spectrum against
the first in the series (masking out low SNR regions). For SSSPM\,0828, the
standard deviation in the shifts was about 0.4 pixels. 
As the wavelength calibration is only accurate to 0.2 pixels, and because the
spectra will anyway be binned (see below) I decided not to apply zero point
offsets.\footnote{Reductions of SSSPM\,0828 with and without these zero point
  wavelength shifts were compared: There was no
  qualitative difference in the results.}

A sky spectrum at the position of the target and reference stars is extracted
directly from the raw (not flat fielded or differenced) images, using the
extraction apertures already defined. These are used together with the
extracted spectra, readout noise estimates and the assumption of Poisson noise
from source and sky to provide an estimate of the noise at each wavelength (the {\em sigma
  spectrum}). This is used only to calculate the \chisq\ spectra.

\subsubsection{Flat fielding}

Flat fielding of the difference images was performed to correct for both
pixel-to-pixel sensitivity variations and global
illumination variations. A global correction is necessary {\em in
  the spatial direction} because the stars are observed
in two different dither positions.  Two types of flat field calibrations were
obtained: (1) Internal flats, using a lamp inside the instrument. The light
path is very short (so there is little atmospheric absorption), but the
illumination is quite different from the telescope pupil illumination (i.e.\ 
it's not ``flat''). (2) External flats using a flat-field screen in the dome,
which have the inverted problem (appropriate pupil illumination but long path
length). The internal flats provide the appropriate high frequency correction
and the external flats the low frequency one. I combine these by
first removing the global variation from the internal flat by dividing it by
its 2D cubic spline fit (with five knots in each dimension) and then multiplying
the result by the same order fit to the external flat (then normalize).
(A similar procedure was used by Bailer-Jones \& Mundt~\cite{bjm01}
for optical imaging.)

It is not possible to achieve an accurate flat field correction in the
dispersion direction, because of the broad absorption features
in the external flat (precluding an accurate definition of the global
fit). However, because the sources are not displaced in the dispersion
direction between the two dither positions, and because we are only using
these data to look for temporal variations in the spectra, an accurate
correction is not required.


\subsection{Spectral rectification, pairing and binning}\label{rectification}

A {\em relative spectrum} is the target spectrum divided by the reference
spectrum. On the assumption that the reference star shows no variations on the
timescales of interest, this should remove variations due to
the Earth's atmosphere.  There were, however, still variations with time of
the integrated flux of the relative spectra.  A similar problem was found in BJ02,
where it was concluded the problem was differential light losses at the slit.  I remove this
by dividing from each relative spectrum its integrated flux
(``rectification''): this puts all relative spectra on a common scale. Because
the full spectral range includes the deep telluric absorption band between the
J and H bands, the flux was integrated only over the regions
0.990--1.316\,\micron\ and 1.489--1.629\,\micron.

Some analyses are later performed on the non-relative spectra of the target or
reference spectra (also rectified by division of the integrated flux). These
will be referred to as {\em direct spectra}.

Plotting the flux vs.\ time for a given wavelength bin showed, for some bins,
a small scale ``zig-zag'' variation in flux between consecutive spectra (i.e.\ 
the two dither positions).  This may be a residual flat fielding problem (or
other detector issue).  This is removed by simply averaging the two spectra
formed at the two consecutive dither positions to give the
{\em paired spectrum}.\footnote{Dithering the telescope was performed in order
  to improve sky subtraction. Given the difficulties of flat fielding to
  sufficient precision and the possibility of other detector variations it is
  not clear that dithering bestowed a net advantage. Given a telescope with
  excellent guiding and observations performed under stable weather conditions, it is
  probably better to observe bright targets without dithering at all.}
All spectra referred to from now on are rectified, paired spectra.

The SNR per pixel in a relative spectrum of SSSPM\,0828 is typically 17
(ranging from 13--20 outside of the strong water absorption band).  This is
increased (at the expense of spectral resolution) by binning the spectra,
specifically, by taking the median of successive blocks of $B$ pixels.  Taking
the median rather than the mean almost entirely removes cosmic rays and bad
pixels.
Note that I {\em bin} rather than {\em smooth} in order that the binned pixels
be independent.  All spectral plots in this paper are for $B=1$ (i.e.\ 
0.696\,nm/pix) and the corelation analysis is done at $B=20$.

\section{Analysis}\label{analysis}

Most previous UCD variability analyses have used just a single
photometric band; the presence of variability is established using a
statistic (typically \chisq\ or the F-statistic) which compares the observed
variations to those expected by the null hypothesis (i.e.\ the noise).
(Koen~\cite{koen04} discusses other statistics.)  This is repeated here, but
now using all of the 1028 bands in the unbinned spectra. The main contribution
of the present work, however, is to look for {\em correlated} variations
between spectral bins, independently of estimated errors. This is done on the
binned spectra (to improve the SNR), with attention paid in particular to known
chemical features or potentially sensitive indices. The assumption is that
some intrinsic variability will be coherent across different parts of the spectrum.
The method is outlined using SSSPM\,0828 and then the results for each target
summarized in section~\ref{results}.

\subsection{Spectra}\label{spectra}

\begin{figure*}
\begin{minipage}{150mm}
  \centerline{
    \includegraphics[width=0.82\textwidth,angle=270]{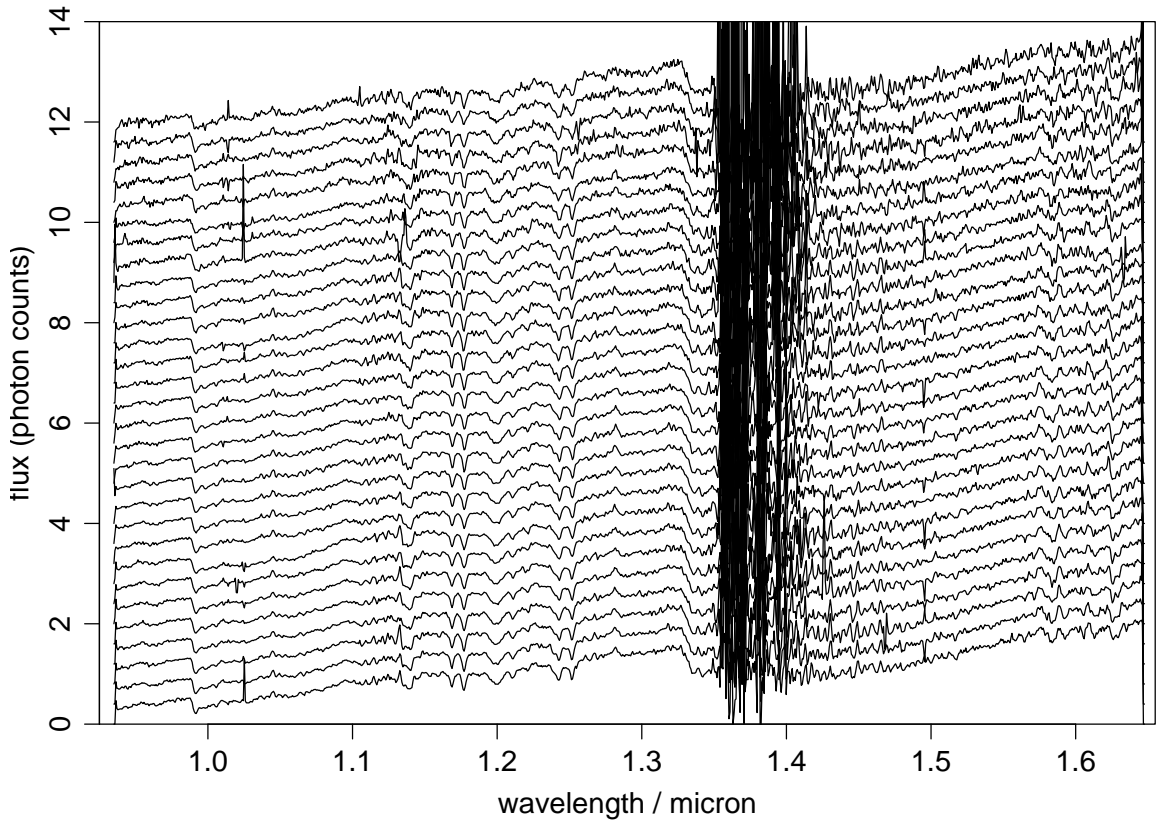}
  }
\caption{The time series of the (unbinned) relative spectra for SSSPM\,0828. Time
    increases up the plot and the spectra have been offset. The spacing
    between consecutive spectra is typically 7.8\,min. Cosmic rays and bad
    pixels have not been removed here.}
\label{ss0828_timeseries}
\end{minipage}
\end{figure*}

The series of 30 relative spectra in the time series of SSSPM\,0828 is shown
in Fig.~\ref{ss0828_timeseries}.\footnote{Movies of the time series are
  available with the online edition of this article. The data themselves are
  available from the author upon request.} The median target (direct) and
relative spectra are shown in Fig.~\ref{ss0828_spectra} (panels A and B).
Panel (C) shows the amplitude of the variations in the form of the median
absolute deviations (MAD) and panel (D) the \chisq\ spectrum, both for the
relative spectrum.  The MAD is a more robust version of the standard
deviation, and includes a correction factor to achieve asymptotic consistency
with the standard deviation of a normal distribution.  The \chisq\ spectrum
(see section~\ref{imagered}) identifies pixels with a large measured variation
compared to the variation predicted from just the noise model.  Thus the
saturated telluric water absorption band between 1.31 and 1.49\,\micron\ (and
to a lesser extent the weaker one at 1.10--1.16\,\micron) shows a large MAD in
Fig.~\ref{ss0828_spectra}, yet a low \chisq\ because the estimated noise is
also large.

Fig.~\ref{zJ_spectra} shows the 0.95--1.32\,\micron\ region of the median
relative spectra for all four targets.

\subsection{Correlation matrices}\label{cor_matrices}

\subsubsection{Definition}

Conclusions drawn from the \chisq\ spectrum alone are sensitive to the estimated errors. As 
discussed by Bailer-Jones \& Mundt (\cite{bjm99}), these are
difficult to estimate accurately, because the basic prescription based on photon
and readout noise neglects other sources. 
I overcome this limitation by looking for variability correlations in
wavelength space. I
calculate the (Pearson) correlation coefficients between all pairs of
(binned) pixels across the whole time series. These are plotted as a matrix in
Fig.~\ref{ss0828_cor_rect_bin20_pair_relspec}, with
the colour scale indicating the modulus of the correlation coefficients,
\modcor. 1.0 indicates perfect correlation (or anti-correlation) and 0.0 no
correlation. The matrix is of course symmetric with unit values on the leading
diagonal.  By plotting \modcor\ rather than \cor\ the plots do not distinguish
between correlations and anti-correlations, which eases comparison with the
matrix for random spectra (see below).  We will see from the histograms in the
next section that all targets have a bias towards stronger positive
correlations, although SDSS\,0539 also shows significant anti-correlations.  We
clearly see structure in the matrix, with some regions of high correlation.
The correlation matrix at higher and lower binning factors shows similar
patterns.  The presence of structure suggests correlated variations in the
UCD. (The issue of autocorrelation is discussed in section~\ref{autocorrelation}).

\subsubsection{Control sample: random spectra}

To determine whether the correlations are statistically significant we require
a control. For this I generate a series of $S$ random spectra in which each
pixel is drawn at random from a $\mathcal{N}(0,1)$ distribution, where $S$ is
the number of spectra in the series ($S$=30 for SSSPM\,0828). An example of
one such correlation matrix is shown in
Fig.~\ref{ss0828_cor_bin20_pair_rnormspec}. Note that its properties are
independent of the mean and standard deviation used. Comparing this
with Fig.~\ref{ss0828_cor_rect_bin20_pair_relspec} we see that the latter
shows many more cells of significant correlation. The same qualitative conclusion can be
drawn from the frequency distribution (histograms) of these correlation
coefficients (Fig.~\ref{ss0828_cor_bin20_pair_histogram}):  The
distribution for the UCD spectra is significantly broader and has
more positive correlations than that for the random spectra.
  
\subsubsection{Statistical significance of correlations}

We can quantify the significance of variations from these control spectra and
their correlation matrices. By looking at the (cumulative) distribution of the
correlation coefficients for large numbers of random spectra we can derive
confidence intervals (this could also be done analytically).

The expectation value of \cor\ is of course zero.  The cumulative
distribution, $C$(\cor), of the correlation coefficients is the probability
that \cor\ is less than some value under the null hypothesis of Gaussian
random variations.  
We are interested in significant
positive or negative correlations, so I make a two-sided test at the 0.1\%
confidence level. As the standard deviation of \cor\ 
(Fig.~\ref{ss0828_cor_bin20_pair_histogram}, bottom) is smaller for larger $S$,
this is done separately for each target.
A set of $S$ spectra is generated and the confidence interval determined. This
is repeated many times (ten thousand) and the average confidence interval
determined (a limiting cumulative distribution is shown in
Fig.~\ref{ss0828_cum_cor_bin20_pair_rnormspec}).
The 0.1\% confidence limits are
(see Table~\ref{obslog}):
$\pm 0.72$ ($S=16$); 
$\pm 0.62$ ($S=25$); 
$\pm 0.61$ ($S=26$); 
$\pm 0.57$ ($S=30$).
Under the assumptions adopted,
values of \modcor\ larger than these indicate significant correlations.

These confidence limits are based on Gaussian
distributions (the random spectra). The variations in the
observed spectra may not be Gaussian, but in the absence of a better model
this is a useful baseline. As a check I also calculated the correlation matrices
using the non-parametric Spearman's rank correlation coefficient.  The
matrices showed very similar patterns, albeit with slightly smaller absolute
values.  The advantage of this correlation analysis over much previous work on
UCD variability is that it makes no assumptions about the values of the
parameters of the noise model. In contrast, the \chisq\ statistic must assume
an amplitude for the noise level, and compare observed variations to this to
test for significance.

\subsubsection{Autocorrelation}\label{autocorrelation}

In principle autocorrelation in the time series of a single (binned) pixel
could produce spurious cross correlations between pixels (e.g.\ Chatfield
\cite{chatfield96}). However, it turns out that the level of autocorrelation
is rather low:  Inspection of the correlograms shows that
the degree of autocorrelation is no higher (or lower!)  than that seen in the
control spectra. This is the case for the full range of time lags. Even in the
region around 1.05\,\micron\ in SSSPM\,0828
(Fig.~\ref{ss0828_cor_rect_bin20_pair_relspec}), for example, where we see
large cross correlation coefficients, these pixels show a low level of
autocorrelation. Both the control spectra and the brown dwarfs spectra can
show non-zero autocorrelation because noise can correlate by chance. This is
taken into account when defining confidence intervals based on the control spectra.

\subsubsection{Significantly correlated regions}

With a binning factor of $B=20$, there are $N=51$ independent binned pixels
and thus $N \times (N-1)/2=1275$ correlation coefficients. With the 0.1\%
confidence intervals we would expect one significant value just by chance.
Partly for this reason I adopt a conservative margin on top of the formal
confidence limits derived above, and define significantly correlated regions
to be those with \modcor\,$>$\,0.8. Table~\ref{correlations} lists those
wavelength bins which show a significant correlation with at least one other
bin: For each spectrum, all bins are listed which have \modcor\,$>$\,0.8 in
the binned relative spectra.  2MASS\,0847 has no significantly correlated
bins. Several bins show a significant correlation with more than one other
bin, as can be seen from the total number of correlated pairs listed at the
bottom of the table.  The binned spectra have 51 binned pixels, and there are
39 distinct entries in the table. That is, 12 spectral bins do not have significant
correlations in any spectrum.  The column ''feature'' lists atomic and
molecular features identified in L and T dwarfs which are coincident with the
given bins. Many of these have been taken from McLean et al.~\cite{mclean03}
and Cushing et al.~\cite{cushing05}. There are some additional, weaker
features which have not been listed. Many of these cannot be seen at the
resolution used in the present work (although unresolved lines/bands could of
course still be variable). For example, Cushing et al.\ identified many FeH
features between 0.998 and 1.085\,\micron, and methane absorption extends over
1.15--1.25\,\micron.

I discuss these significant correlations for
each object in section~\ref{results}.

\begin{table}
\caption[]{
Wavelength bins (in nm) which show a significant correlation (\modcor\,$>$\,0.8) with at
least one other bin.  
Those in bold have \modcor\,$>$\,0.9. This is obtained for a B=20 binning
where the bin width is 13.2\,nm.
The column ''feature'' lists atomic and molecular features identified in L 
and T dwarf spectra (although not
necessarily present at all spectral types) which are coincident with the given
bins and thus potentially responsible for the detected correlated variability. Also indicated in this column
are telluric OH lines.
A question mark (``?'')
indicates no specific feature. The total number of
{\em pairs} of bins with \modcor\,$>$\,0.8 is given in the penultimate table row and
the number with \modcor\,$>$\,0.9 is in the final row. 
\label{correlations}
}
\begin{tabular}{lrrrr}
\hline
Feature & SDSS & 2MASS & SSSPM & 2MASS \\
        & 0539 &  0559 &  0828 &  0847 \\
\hline
\h2o    &               &        942    &               &               \\
\h2o, VO &               &        955    &               &               \\
\h2o    &        969    &               &               &               \\
FeH \h2o        &       {\bf 983}       &               &               &               \\
Fe FeH CrH      &        997    &               &        997    &               \\
?       &               &               &       {\bf 1011}      &               \\
?       &       1025    &               &       {\bf 1025}      &               \\
?       &       1039    &               &       {\bf 1039}      &               \\
VO      &       {\bf 1053}      &               &       {\bf 1053}      &               \\
?       &       1066    &               &       {\bf 1066}      &               \\
? OH    &       {\bf 1080}      &               &       1080    &               \\
TiO OH  &       {\bf 1094}      &               &               &               \\
?       &       1108    &               &       1108    &               \\
\h2o    &               &               &       1122    &               \\
\k1     &       1177    &               &               &               \\
Fe FeH  &       1191    &               &               &               \\
FeH     &       1205    &               &               &               \\
FeH OH  &       1219    &               &               &               \\
\k1     &       1247    &               &               &               \\
?       &       {\bf 1261}      &               &               &               \\
?       &       {\bf 1275}      &               &       1275    &               \\
\ch4 \ti1 OH    &       {\bf 1288}      &               &       1288    &               \\
\ch4 \h2o OH    &       {\bf 1302}      &               &       1302    &               \\
\ch4 \ca1 \al1  &       {\bf 1316}      &               &               &               \\
\h2o    &               &       1344    &               &               \\
\h2o    &               &       1441    &               &               \\
\h2o    &       {\bf 1469}      &               &               &               \\
\h2o    &       1483    &               &               &               \\
\h2o    &       {\bf 1497}      &               &               &               \\
\k1     &       1511    &               &       1511    &               \\
\k1 OH  &       {\bf 1524}      &               &       1524    &               \\
?       &       1538    &               &       1538    &               \\
?       &       {\bf 1552}      &               &       1552    &               \\
?       &       {\bf 1566}      &               &       1566    &               \\
FeH OH  &       {\bf 1580}      &               &               &               \\
FeH     &       {\bf 1594}      &               &               &               \\
\ch4 OH &       {\bf 1608}      &               &               &               \\
\ch4 OH &       1622    &               &               &               \\
\ch4    &       {\bf 1635}      &               &               &               \\
        &               &               &               &               \\
N(\modcor\,$>$\,0.8)    &       114     &       4       &       23      &       0       \\
N(\modcor\,$>$\,0.9)    &       {\bf 19}        &       {\bf 0} &       {\bf 10}        &       {\bf 0} \\
\hline
\end{tabular}
\end{table}

\subsubsection{Correlations are not due to OH lines}

An imperfect sky subtraction could, in principle, lead to variability in the
OH lines, and to examine this the stronger lines are also listed in
Table~\ref{correlations}.  While some of the variable regions coincide with OH
lines, many bins at the location of OH lines are not variable.
Quantitatively, of the 20 strong OH lines present in the observed wavelength
region, 11 are at wavelengths which show no variability in any of the observed
UCDs (so are not listed) and these are not simply the weaker OH lines.
Furthermore, none of the correlated bands in 2MASS\,0559 and none of the more
significantly (\modcor\,$>$\,0.9) correlated bands in SSSPM\,0828 are associated
with OH features. SDSS\,0539 has quite a few correlated bins, but of those listed
in Table~\ref{correlations} which correspond to OH lines, the bins they
correlate with mostly do {\em not} contain OH lines: Of the 19 pairs with
\modcor\,$>$\,0.9 only two have OH lines in both bins.  In conclusion, while we
cannot rule out telluric OH lines as causing {\em some} apparent variability,
it is certainly not responsible for a significant fraction of it.

\section{Results}\label{results}

\subsection{SSSPM J0829-1309 (SSSPM\,0828)}\label{SS0828}

\begin{figure*}
\begin{minipage}{150mm}
\centerline{
\includegraphics[width=0.82\textwidth,angle=270]{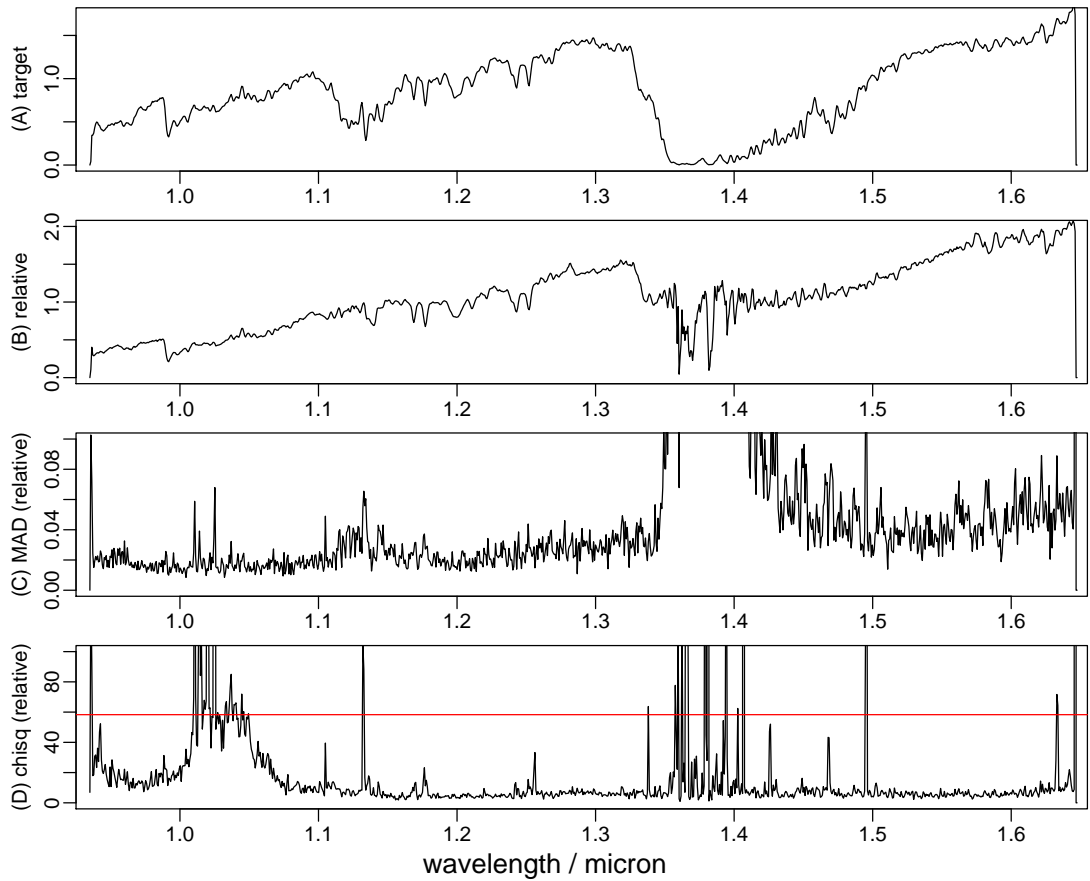}
}
\caption{Spectra of the L2 dwarf SSSPM\,0828 (unbinned at a dispersion of
  0.696\,nm/pix, some 1028 pixels). The flux scale is proportional to photon
  counts (not energy). From top to bottom: (A) The median target spectrum; (B)
  the median relative spectrum (median of target divided by reference at each
  epoch). Because both are rectified by dividing by the integrated flux (see
  section~\ref{rectification}) the vertical scale is dimensionless. (C) The
  median absolute deviation in the relative spectrum (a robust version of the
  standard deviation); (D) The \chisq\ spectrum. In this bottom panel, points
  above the horizontal line are variations beyond the estimated errors with a
  confidence of 99.9\% or more per pixel.}
\label{ss0828_spectra}
\end{minipage}
\end{figure*}

\begin{figure*}
\begin{minipage}{150mm}
  \centerline{
    \includegraphics[width=0.70\textwidth,angle=270]{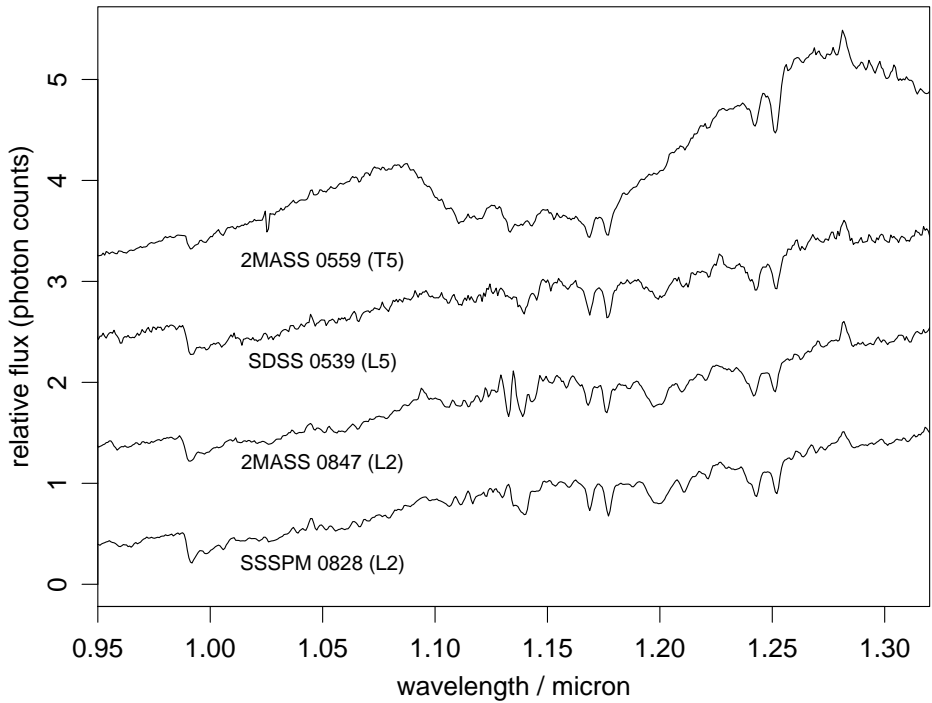}
  }
\caption{Comparison of the median relative spectra. The spectra have been
    offset by values of 0,1,2,3 in the vertical scale (these numbers being the
    zero photon count level for each spectrum). See section~\ref{results}, in
    particular section~\ref{SS0828}, for a discussion of the features.}
\label{zJ_spectra}
\end{minipage}
\end{figure*}

The direct and relative spectra of the L2 dwarf SSSPM\,0828 in
Fig.~\ref{ss0828_spectra} clearly show the presence of the \k1\ doublets at
1169/1178\,nm and 1244/1253\,nm as well as the FeH Wing-Ford band at
990--994\,nm (McLean et al.~\cite{mclean03}). A zoom of this region is shown
in Fig.~\ref{zJ_spectra}. The absorption line at 1140\,nm is probably the
unresolved \na1\ doublet at 1138/1141\,nm.  We see a large absorption feature
centered at 1199\,nm, which is most likely due to the FeH band head at
1194\,nm, possibly blended with an Fe line at 1197\,nm.
There is marginal indication of the \al1\ lines at 1313\,nm (perhaps blended
with \ca1\ at 1314\,nm) and 1315\,nm, but no sign of the 
\mn1\ (1290\,nm) or \rb1\ (1323\,nm) lines. The feature around 1517\,nm is
probably \k1.
Comparing the continua of the target and relative spectra (panels A and B), we
see that, in this case, the absorption band between 1.1 and 1.16\,micron must
be almost entirely due to telluric absorption, because it is not present in
the relative spectrum. This is not the case for the wider band between 1.31 and
1.49\,\micron, however, where we still see significant absorption in the
relative spectrum.  This is clear evidence for the presence of water in this
L2V object.

We see a resolved peak in the spectrum at 1282\,nm. It is present in the
relative spectra of all four targets. This coincides with the wavelength of
the Paschen $\beta$ line, identified by Cushing et al.~\cite{cushing05} in the
spectrum of Arcturus (K1.5) and M dwarfs down to M7.  Inspection of the direct
reference spectra shows an absorption feature at this wavelength. If 
all of the reference stars are K to mid M
dwarfs (not implausible), it would be intrinsic to them, not present in the target spectra, and
therefore would show up as a peak in the relative spectrum.  
Alternatively this peak could be the
``continuum'' between two absorption features: the absorption on the red side
is plausibly caused by \ti1\ at 1285\,nm.  \ti1\ is dominant in its monatomic
form above atmospheric temperatures of 2500\,K, although such layers are below
an optical depth of unity for \teff$>$ 1800 and 2000\,K in the dusty and cond
models (respectively) of Allard et al.~\cite{allard01} (their Figs.~5 and 6).
At the lower temperatures higher in the atmosphere Ti becomes increasingly locked
into TiO, finally being removed as CaTiO$_3$ at 1957\,K (Lodders
\cite{lodders99}).

\begin{figure}
  \centerline{
    \includegraphics[width=0.45\textwidth,angle=270]{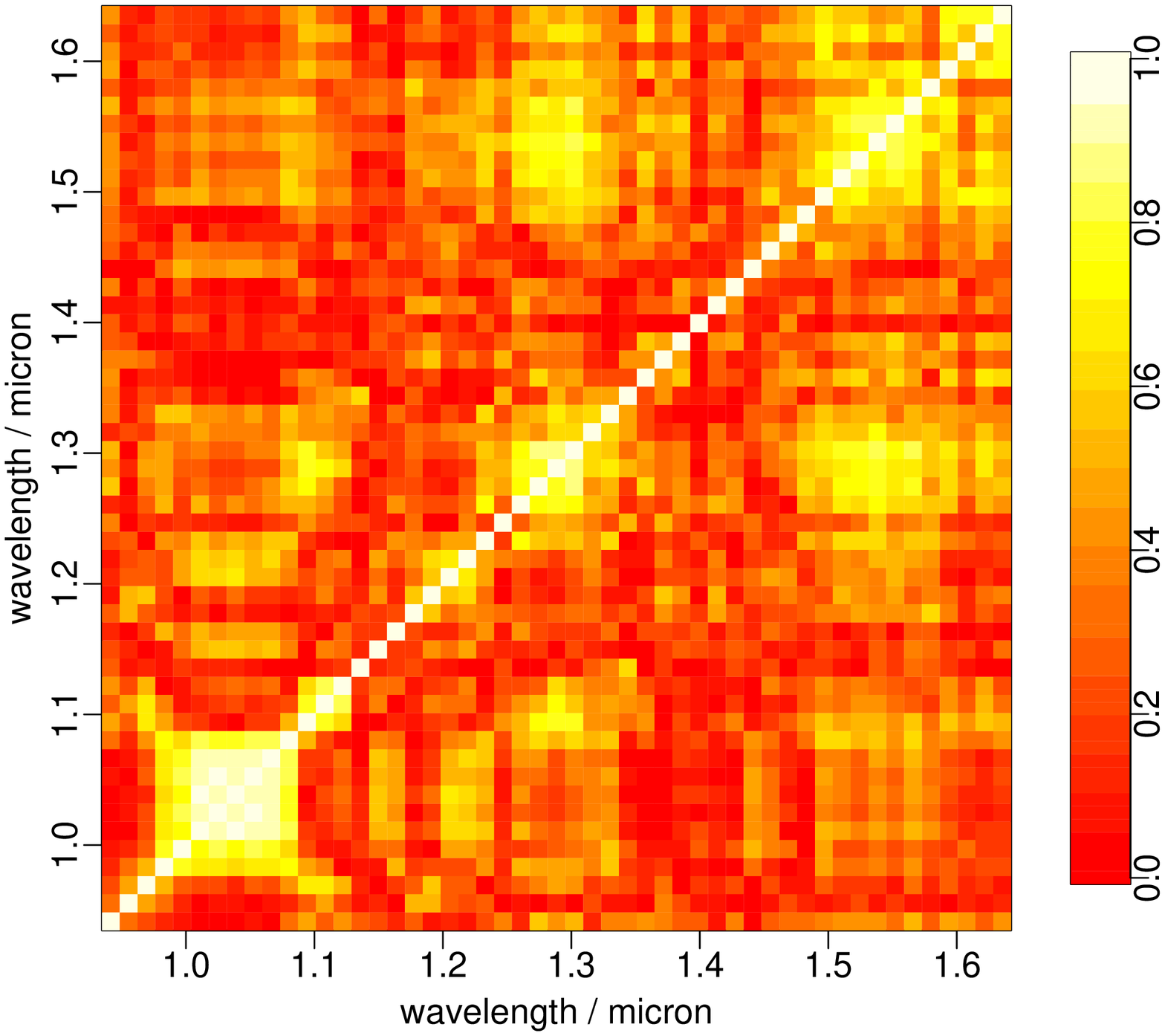}
  }
\caption{Correlation matrix for the relative spectra of SSSPM\,0828 with a binning
    factor of 20. The absolute values of the correlation coefficients are shown.}
\label{ss0828_cor_rect_bin20_pair_relspec}
\end{figure}

The correlation matrix (Fig.~\ref{ss0828_cor_rect_bin20_pair_relspec}) shows
quite a few regions with high correlation, 17 with \modcor\,$>$\,0.8
(Table~\ref{correlations}). Ten different pairings of five binned pixels show
correlations with \modcor\,$>$\,0.9, all lying in the range 1010--1070\,nm. Some
of the \modcor\,$>$\,0.8 pixels are coincident
with known features, including Fe, FeH, VO, \k1\ and water, and some also have
statistically significant \chisq\ values. It is interesting that the 17 bins
are grouped into five contiguous regions. In conclusion, while we have evidence for
significant correlated variability, it is not restricted to
readily-identifiable features. The variability could be caused by dust or
molecules (which would effect the broader ``continuum'' rather than lines),
for example FeH or water (methane we may want to rule out a priori based on
the spectral type).

\begin{figure}
\centerline{
\includegraphics[width=0.45\textwidth,angle=270]{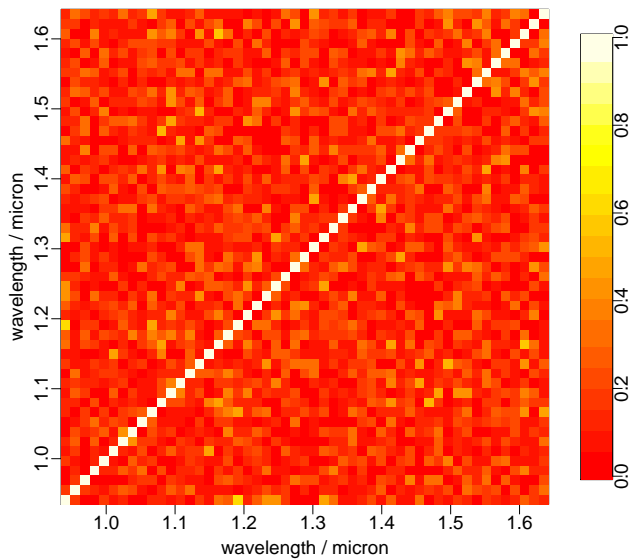}
}
\caption{Correlation matrix for a series of random spectra. The same
  dispersion function (and binning) has been used as for the science
  spectra. This example uses a set of 30 spectra (the same as SSSPM\,0828).
\label{ss0828_cor_bin20_pair_rnormspec}
}
\end{figure}

\begin{figure}
\centerline{
\includegraphics[width=0.75\textwidth,angle=270]{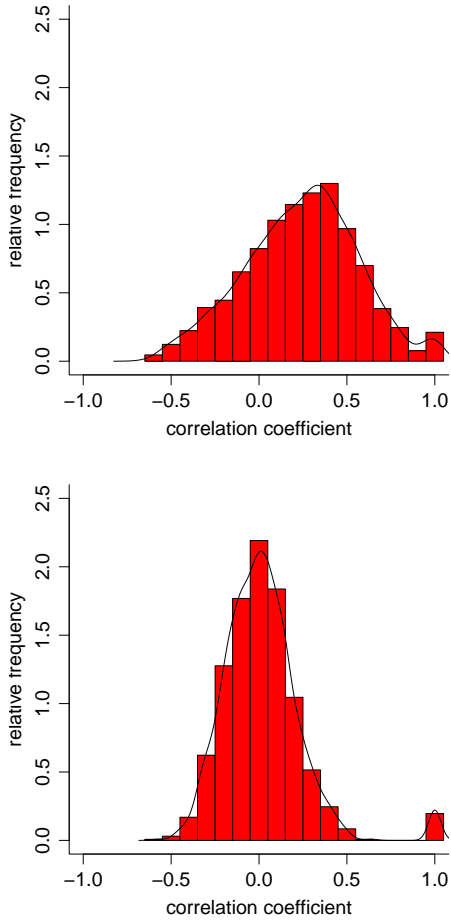}
}
\caption{Histogram of the correlation coefficients for SSSPM\,0828 (top; from
  Fig.~\ref{ss0828_cor_rect_bin20_pair_relspec}) and for the random
  spectra (bottom; from Fig.~\ref{ss0828_cor_bin20_pair_rnormspec}). The ordinate is
  a density (i.e.\ the total area is 1.0). The overplotted line is a kernel
  density estimate.
\label{ss0828_cor_bin20_pair_histogram}
}
\end{figure}

\begin{figure}
\centerline{
\includegraphics[width=0.30\textwidth,angle=270]{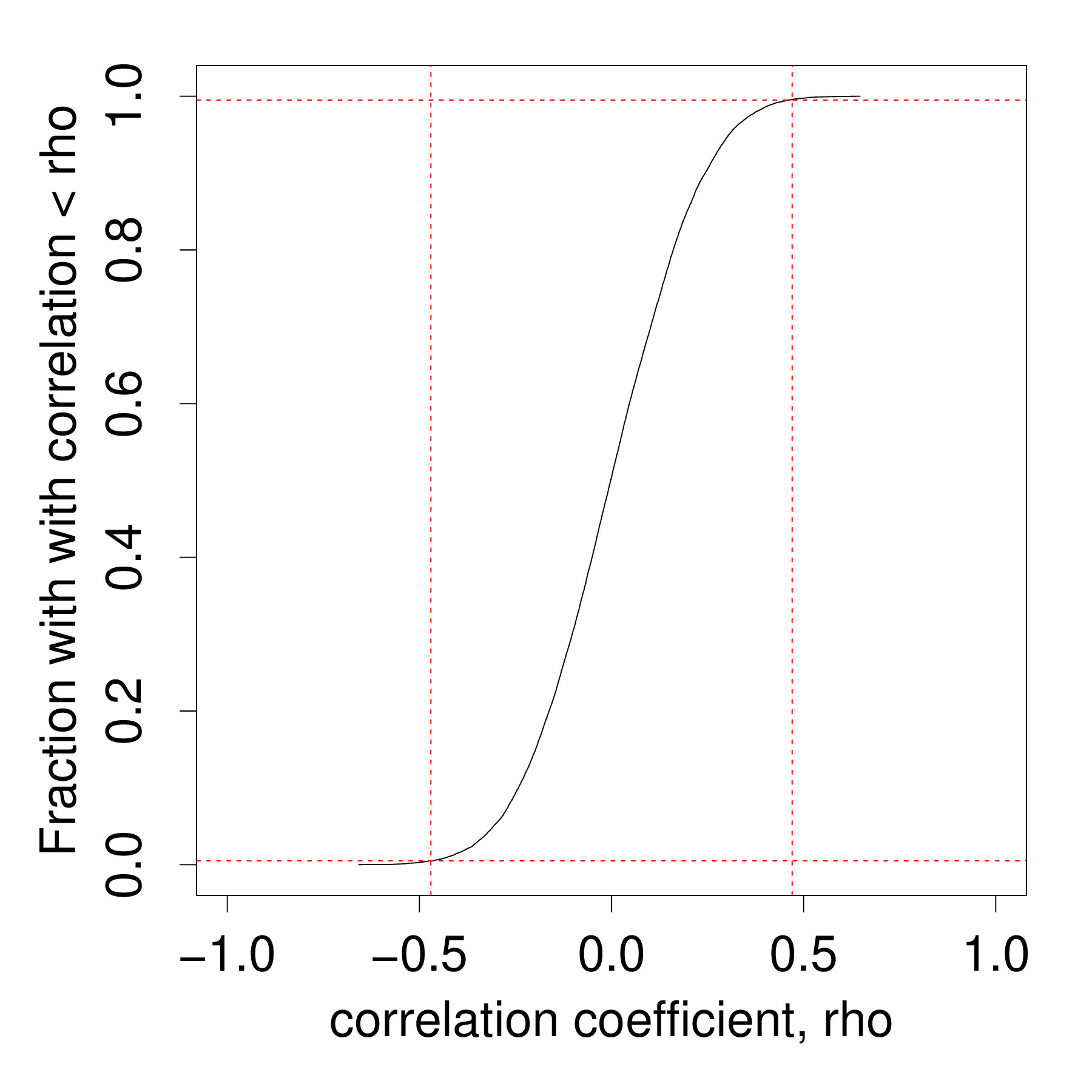}
}
\caption{Limiting cumulative distribution of the correlation coefficients, $C$(\cor), for the random spectra (Fig.~\ref{ss0828_cor_bin20_pair_histogram}, bottom), excluding the leading diagonal of unit correlations. The horizontal lines are at 0.005 and 0.995, i.e.\ there is only a 1\% chance that the correlation coefficient could lie outside these limits for random data. The limits are $\pm$ 0.46 in this case ($S=30$).
\label{ss0828_cum_cor_bin20_pair_rnormspec}
}
\end{figure}

\subsection{2MASS J05591914-1404488 (2MASS\,0559)}

\begin{figure*}
\begin{minipage}{150mm}
\centerline{
\includegraphics[width=0.82\textwidth,angle=270]{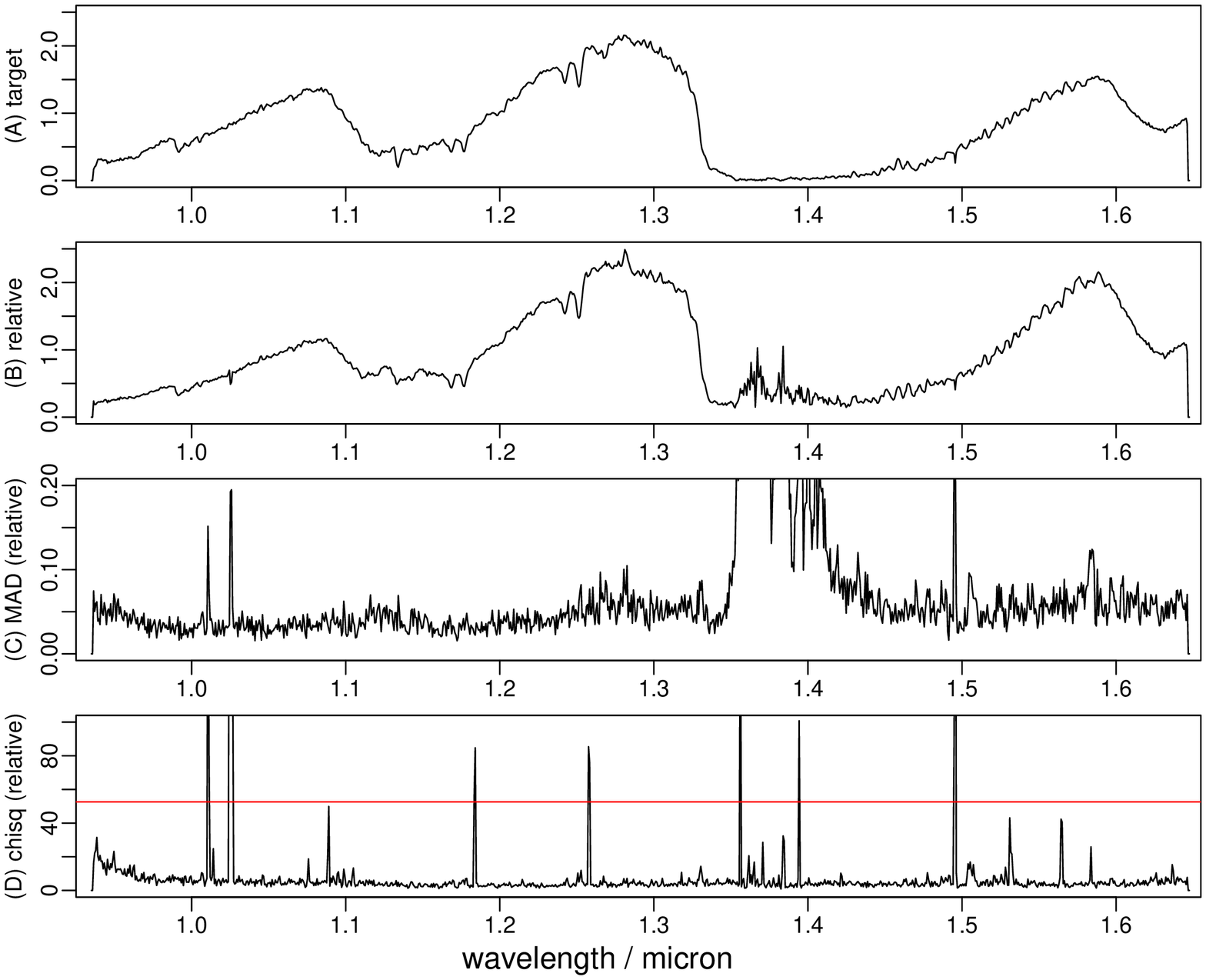}
}
\caption{Spectra of the T5 dwarf 2MASS\,0559. See caption to Fig.~\ref{ss0828_spectra}.
\label{2m0559_spectra}
}
\end{minipage}
\end{figure*}

The target and relative spectra of this T5 dwarf (Fig.~\ref{2m0559_spectra})
show features typical of UCDs, in particular the \k1\ doublets at
1169/1178\,nm and 1244/1253\,nm, as with SSSPM\,0828 (see also
Fig.~\ref{zJ_spectra}).  A feature which may be the \na1\ doublet is barely
visible at 1138/1141\,nm. The deeper feature near to this is actually centered
on 1133\,nm (and there is another broader one at 1143\,nm).  The FeH Wing-Ford
band can clearly be seen at 990--994\,nm.  The putative FeH/Fe feature at
1199\,nm seen in SSSPM\,0828 is barely visible.  We see a strong absorption band
extending redwards from about 1.31\,nm out to 1.5--1.6\,\micron.  It is much
deeper and extends bluer than the absorption in SSSPM\,0828 (an L2).  This is
almost certainly due to water and methane, and as it is seen in both the
target and the relative spectra it is certainly intrinsic to the target. The
``ripples'' superimposed on this may be many weak absorption features. The
absorption band redwards of 1.6\,\micron\ is likely due to methane; it is
entirely lacking in the SSSPM\,0828 spectrum.

We see another deep absorption feature between 1.08 and 1.2\,\micron. It again
extends much further to the red than the similar band in SSSPM\,0828 and could
again be increased water absorption in the atmosphere of 2MASS\,0559.  Note that
just because the telluric water band causes large variations (panel C) between
1.35 and 1.40\,\micron, these variations are not necessarily significant
(panel D), because of the larger predicted noise.  A number of the significant
peaks in the \chisq\ spectrum are due to cosmic rays (e.g.\ 1.089, 1.184 and
1.258\,\micron) or bad/hot pixels (e.g.\ 1.011, 1.025, 1.495\,\micron) because
we see them in just one spectrum or at one of the two detector positions
(respectively).

\begin{figure}
\centerline{
\includegraphics[width=0.45\textwidth,angle=270]{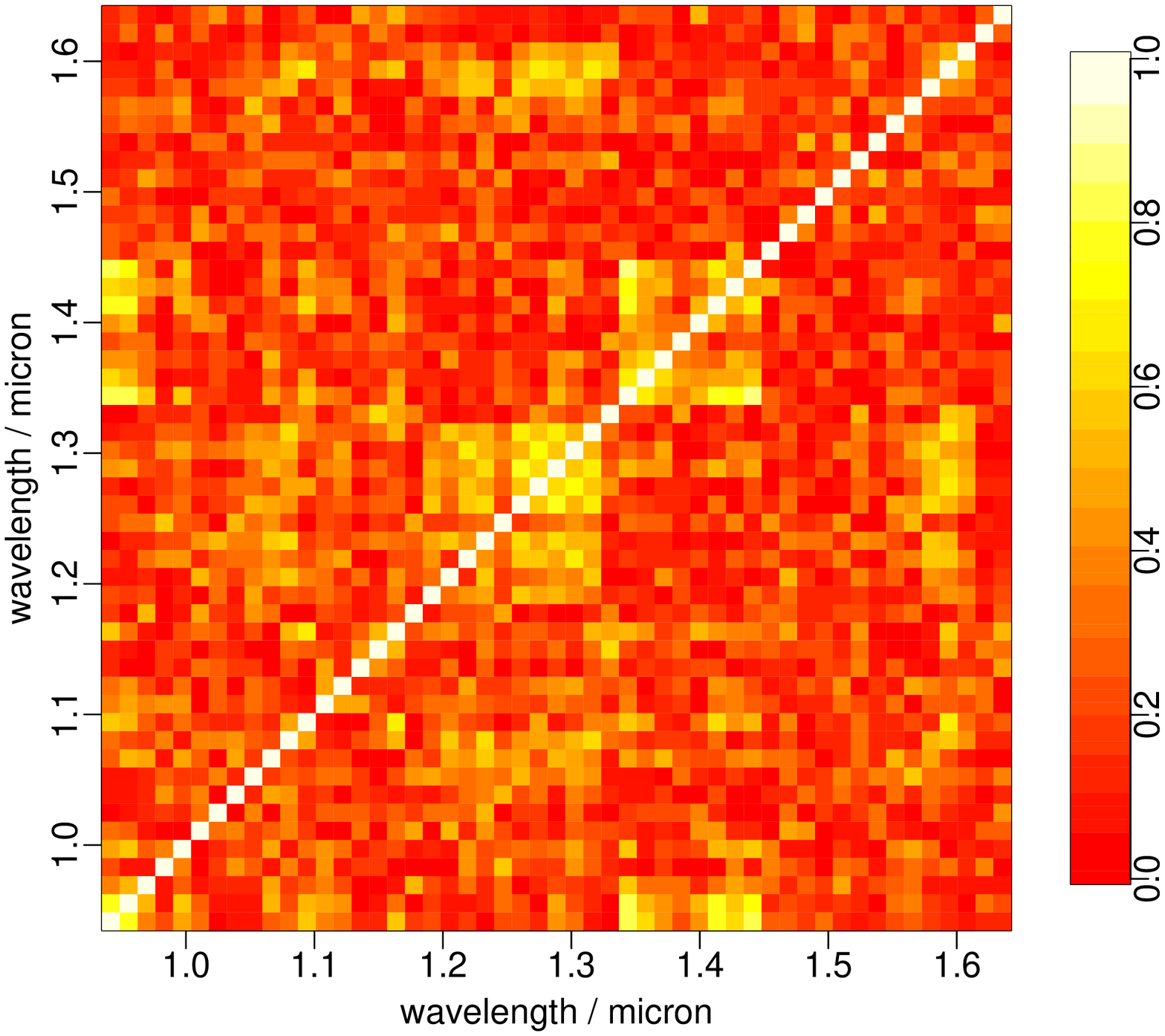}
}
\caption{Correlation matrix for the relative spectra of 2MASS\,0559 with a binning
  factor of 20. The absolute values of the correlation coefficients are shown.}
\label{2m0559_cor_rect_bin20_pair_relspec}
\end{figure}

\begin{figure}
\centerline{
\includegraphics[width=0.4\textwidth,angle=270]{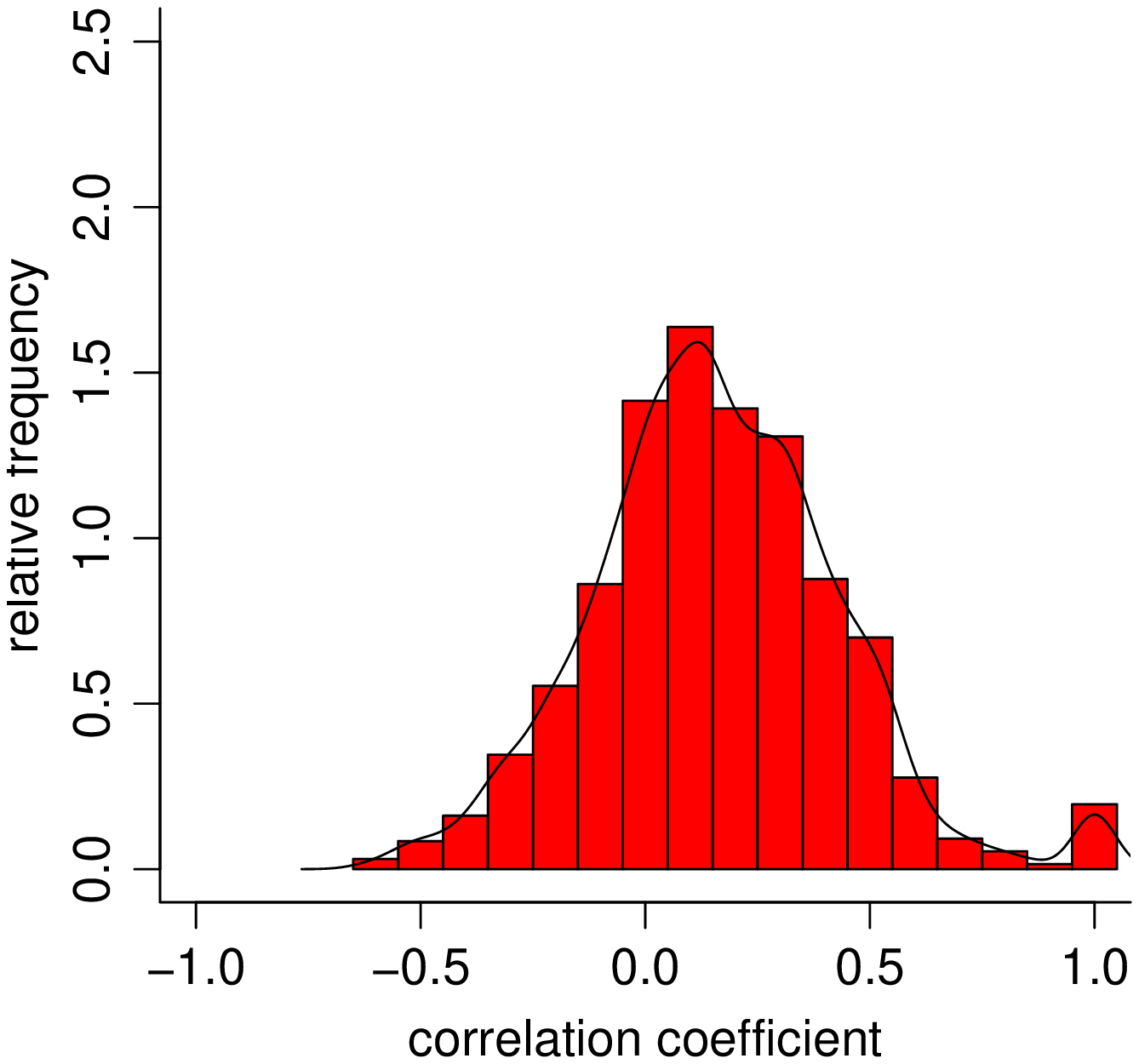}
}
\caption{Histogram of the correlation coefficients for 2MASS\,0559 from Fig.~\ref{2m0559_cor_rect_bin20_pair_relspec}.}
\label{2m0559_cor_bin20_histogram}
\end{figure}

The correlation matrix (Fig.~\ref{2m0559_cor_rect_bin20_pair_relspec}) and
histogram (Fig.~\ref{2m0559_cor_bin20_histogram}) show a few regions of high
correlation (when compared to the random matrix).  Above the threshold
\modcor\,$>$\,0.8 there are four binned pixels (Table~\ref{correlations}), all
of which coincide with water features. This is discussed further in
section~\ref{water_var}. Whether or not the lack of (or lower level of)
correlated variability in other water bands in 2MASS\,0559 is consistent with this
requires a more detailed analysis of the water spectrum.  

\subsection{SDSSp J053951.99-005902.0 (SDSS\,0539)}\label{SD0539}

\begin{figure*}
\begin{minipage}{150mm}
\centerline{
\includegraphics[width=0.82\textwidth,angle=270]{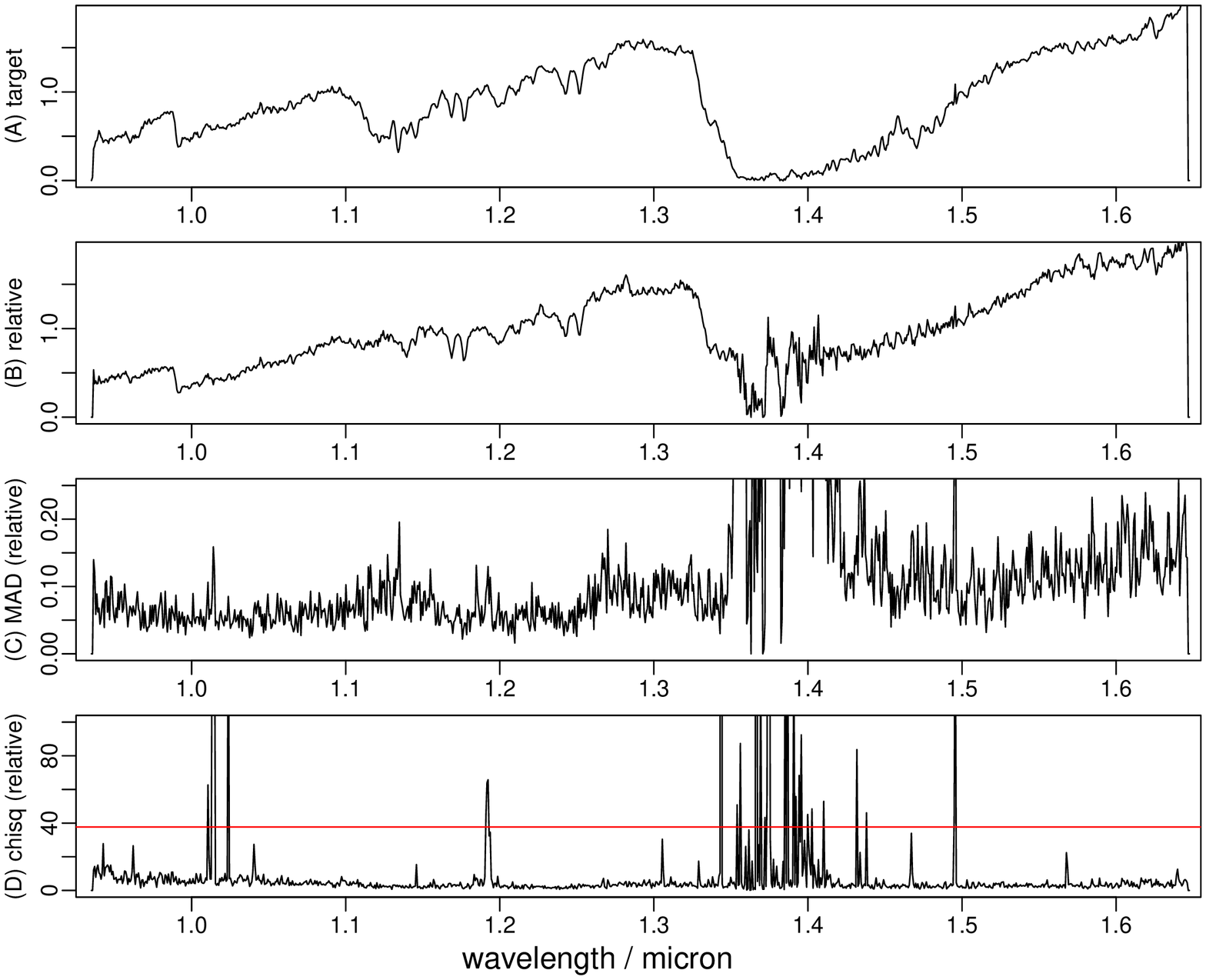}
}
\caption{Spectra of the L5 dwarf SDSS\,0539. See caption to Fig.~\ref{ss0828_spectra}}
\label{sd0539_spectra}
\end{minipage}
\end{figure*}

The spectra for this L5V object are shown in Fig.~\ref{sd0539_spectra}
and Fig.~\ref{zJ_spectra}. 
Although a later type than SSSPM\,0828, many of the features are common, so see
section~\ref{SS0828} for a discussion. In SDSS\,0539 we again see a non-saturated
water band in the relative spectrum. Within this, we see two conspicuous
absorption features, one centered at 1362\,nm and another at 1371\,nm. Similar
features are seen in SSSPM\,0828 at the same position (to within $\pm 1$\,nm). This
may be genuine structure within the water absorption band, although this
region is quite noisy due to low photon counts from the source.

We see that the strength of this absorption band redward of 1.31\,\micron\ is
intermediate between that seen in the two L2 dwarfs (SSSPM\,0828 and 2MASS\,0847) and
the T5 dwarf (2MASS\,0559). As the depth of the latter is due to both water and
methane, this could indicate the presence of some methane absorption in
SDSS\,0539, an L5 object. Methane absorption has been observed in late L dwarfs
before, e.g.\ by Nakajima et al. \cite{nakajima01} in a L6.5V at 1.62 and
1.67\,\micron\ and by Schweitzer et al.  \cite{schweitzer02} at 2.2 and
3.3\,\micron\ (L6--L8), but not at this particular band.

The \chisq\ spectrum shows some evidence for variability. In their
I-band monitoring program, Bailer-Jones \& Mundt~\cite{bjm01} found
SDSS\,0539 to be significantly variable and even detected a significant period at
$13.3\pm1.2$\,hrs.

\begin{figure}
\centerline{
\includegraphics[width=0.45\textwidth,angle=270]{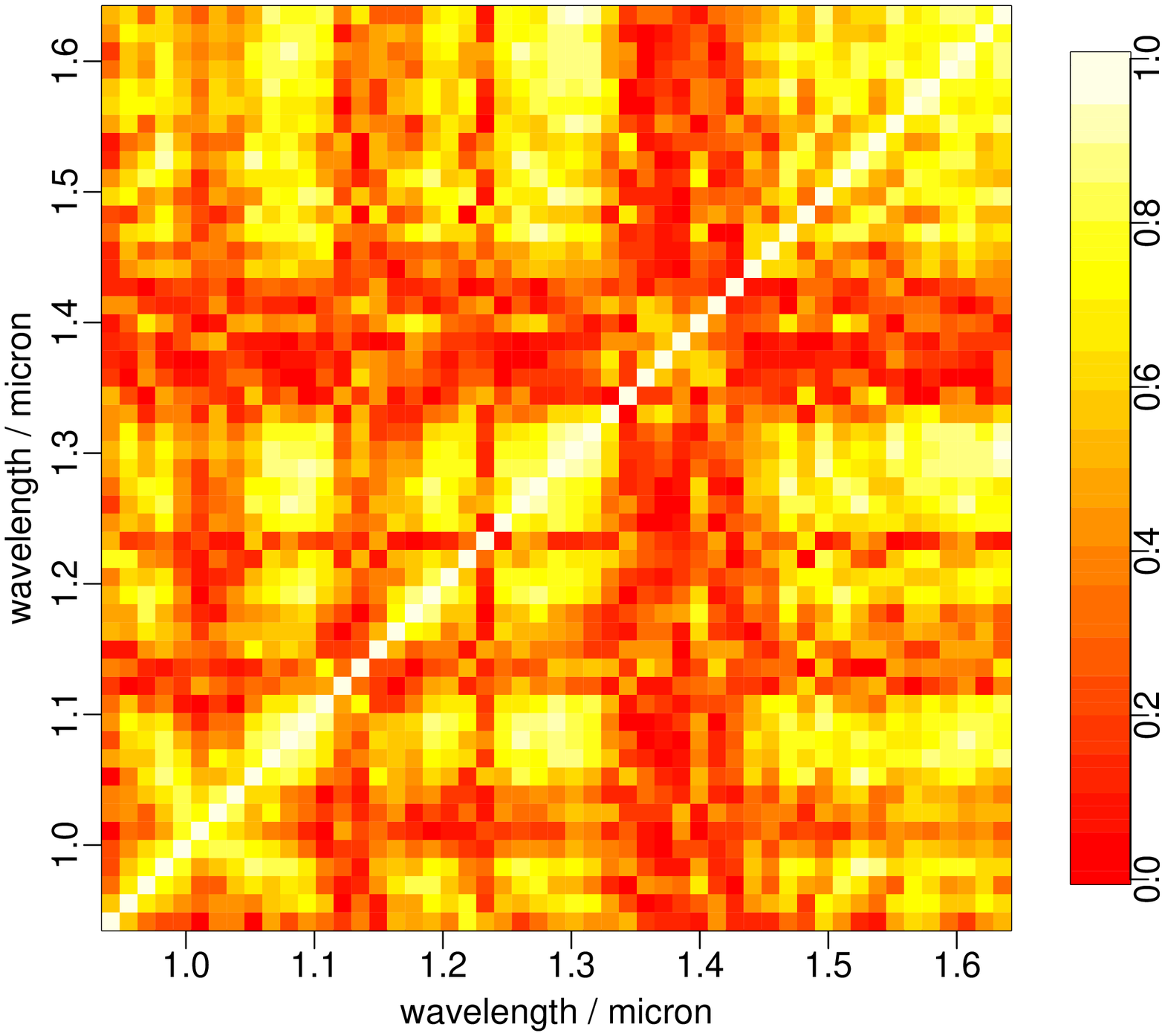}
}
\caption{Correlation matrix for the relative spectra of SDSS\,0539 with a binning
  factor of 20. The absolute values of the correlation coefficients are shown.}
\label{sd0539_cor_rect_bin20_pair_relspec}
\end{figure}

\begin{figure}
\centerline{
\includegraphics[width=0.42\textwidth,angle=270]{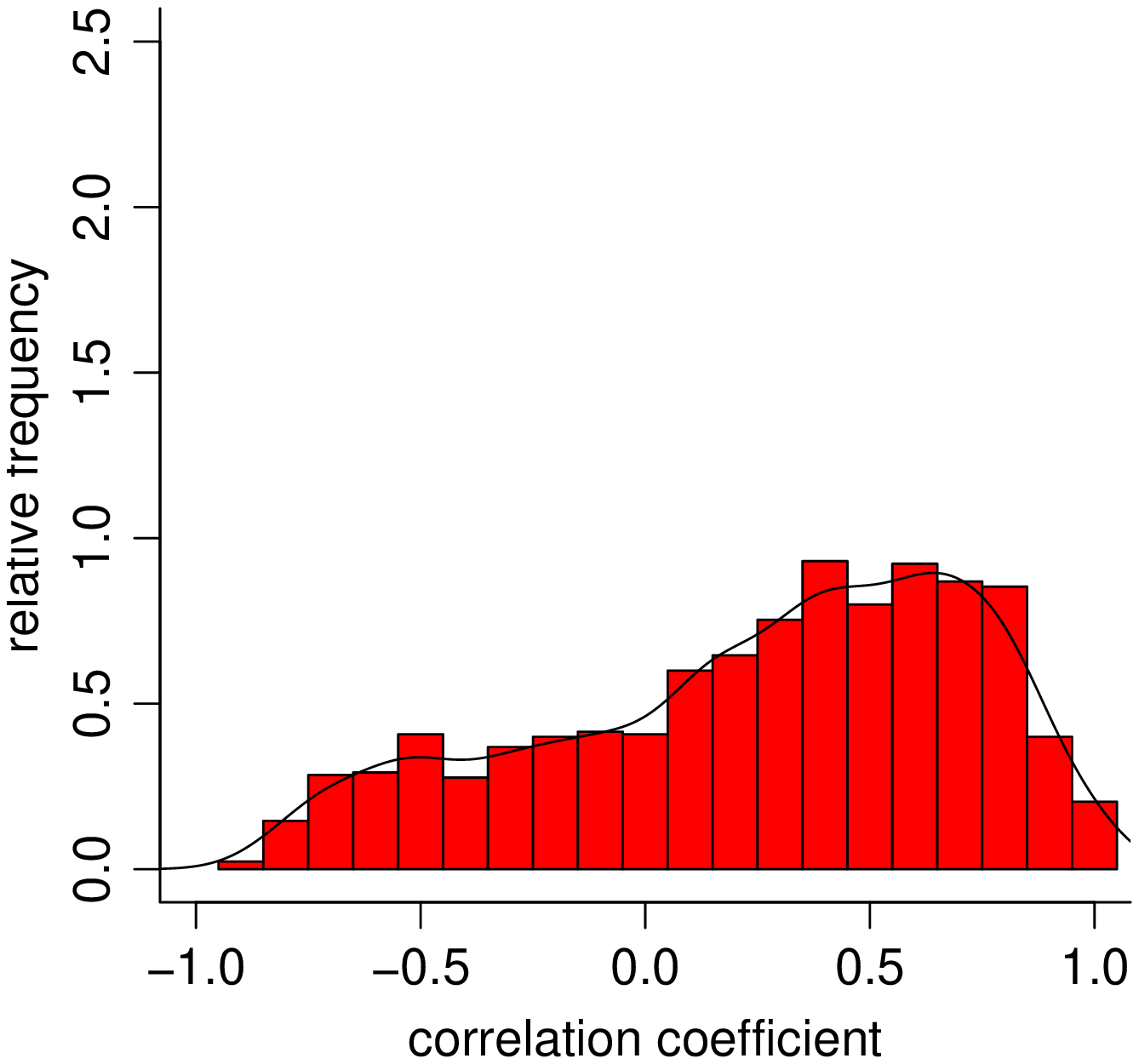}
}
\caption{Histogram of the correlation coefficients for SDSS\,0539 from Fig.~\ref{sd0539_cor_rect_bin20_pair_relspec}.}
\label{sd0539_cor_bin20_histogram}
\end{figure}

The correlation matrix (Fig.~\ref{sd0539_cor_rect_bin20_pair_relspec}) shows,
in comparison to the previous two objects, significantly more highly correlated
binned pixels, as the histogram also makes clear
(Fig.~\ref{sd0539_cor_bin20_histogram}).
33 distinct binned pixels are significantly correlated, 18 of these with 
\modcor\,$>$\,0.9 (see Table~\ref{correlations}).
While the evidence for correlated variability is clear, it is not obviously
connected to a few specific chemical elements. Many of the correlated binned
pixels are contiguous, which points to something with a broad spectral
signature being responsible, such as solid or liquid particulates (``dust'').
Several regions are in common to SSSPM\,0828. One may be a little skeptical with
the extent of variability, given that this is the faintest object and the spectra
are quite noisy.  In theory, lower SNR data does not give rise to more
significant correlations.  On the contrary, the larger noise would dilute any
intrinsic correlations. Moreover, the object with the second most
correlated binned pixels in the sample of four, SSSPM\,0828, is by far the
brightest.

Of the four targets, SDSS\,0539 is the only one to show significant {\em
  anti}correlations, with ten pairs fulfilling the criterion \cor $<-0.8$. Of
these, six involve the bin centered at 1191\,nm, which covers Fe and FeH
features (Table~\ref{correlations}), and this anti-correlates with, amongst
others, bins at 1094\,nm (TiO) and 1219\,nm (FeH). Another pair which can be
associated with known features is 1094\,nm (TiO) and 1205\,nm (FeH).  Of
course, this is not conclusive evidence that these molecular features are the
cause (they sit on top of a continuum, after all). 
But it would be interesting to see whether dynamic atmospheric models show such
anti-correlations.

\subsection{2MASS J08472872-1532372 (2MASS\,0847)}

\begin{figure*}
\begin{minipage}{150mm}
\centerline{
\includegraphics[width=0.82\textwidth,angle=270]{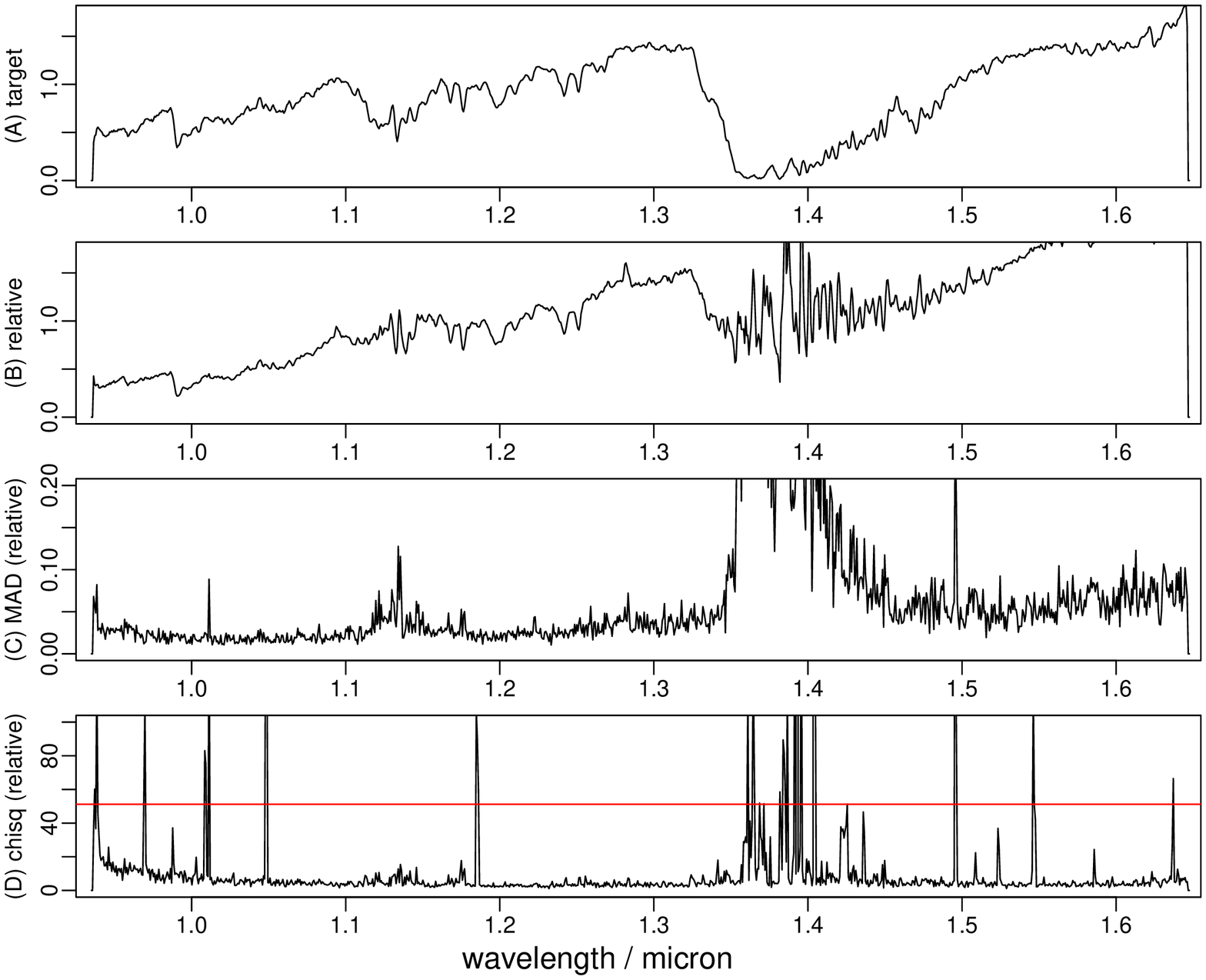}
}
\caption{Spectra of the L2 dwarf 2MASS\,0847. See caption to Fig.~\ref{ss0828_spectra}}
\label{2m0847_spectra}
\end{minipage}
\end{figure*}

\begin{figure}
\centerline{
\includegraphics[width=0.42\textwidth,angle=270]{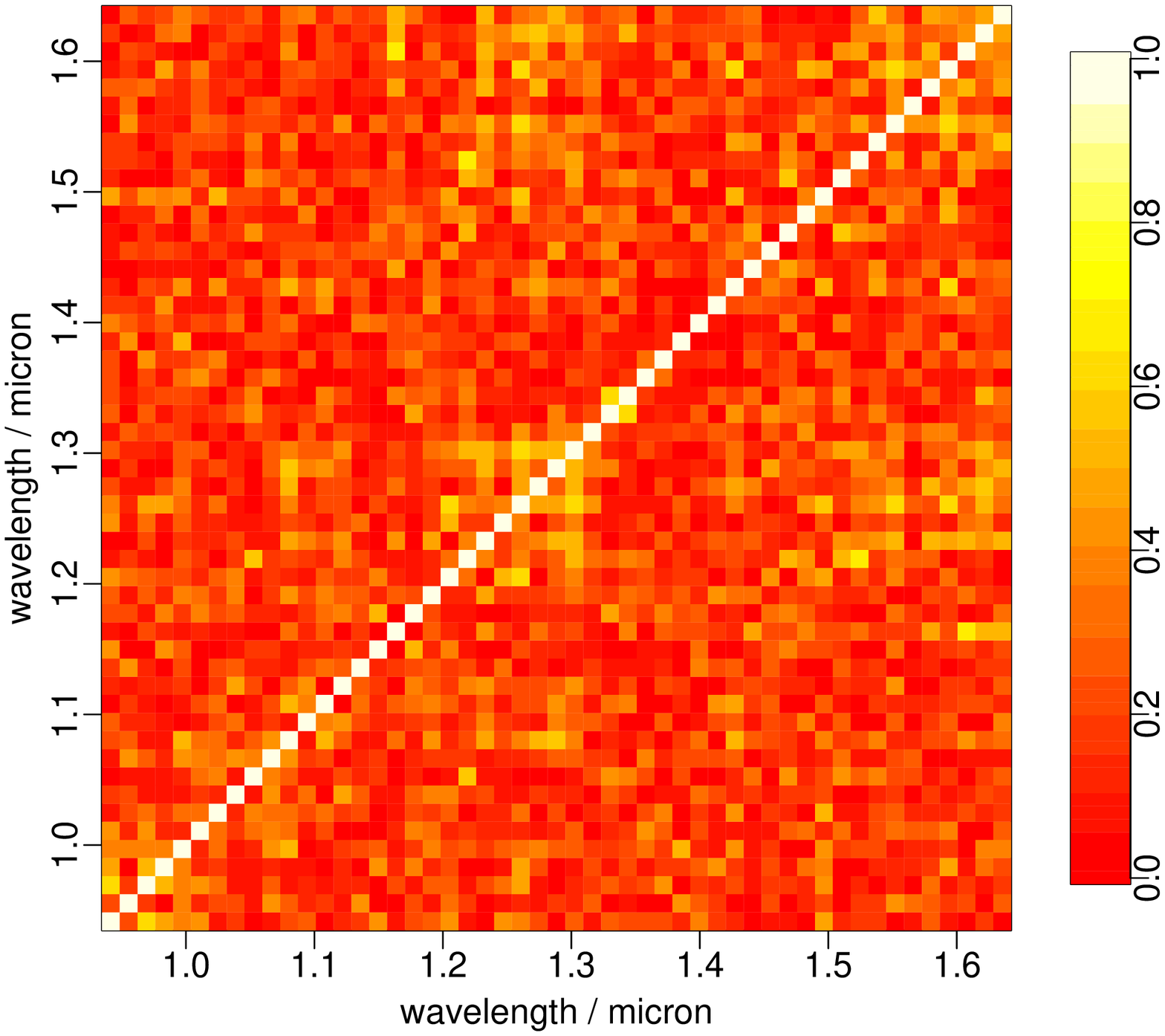}
}
\caption{Correlation matrix for the relative spectra of 2MASS\,0847 with a binning
  factor of 20. The absolute values of the correlation coefficients are shown.}
\label{2m0847_cor_rect_bin20_pair_relspec}
\end{figure}

\begin{figure}
\centerline{
\includegraphics[width=0.42\textwidth,angle=270]{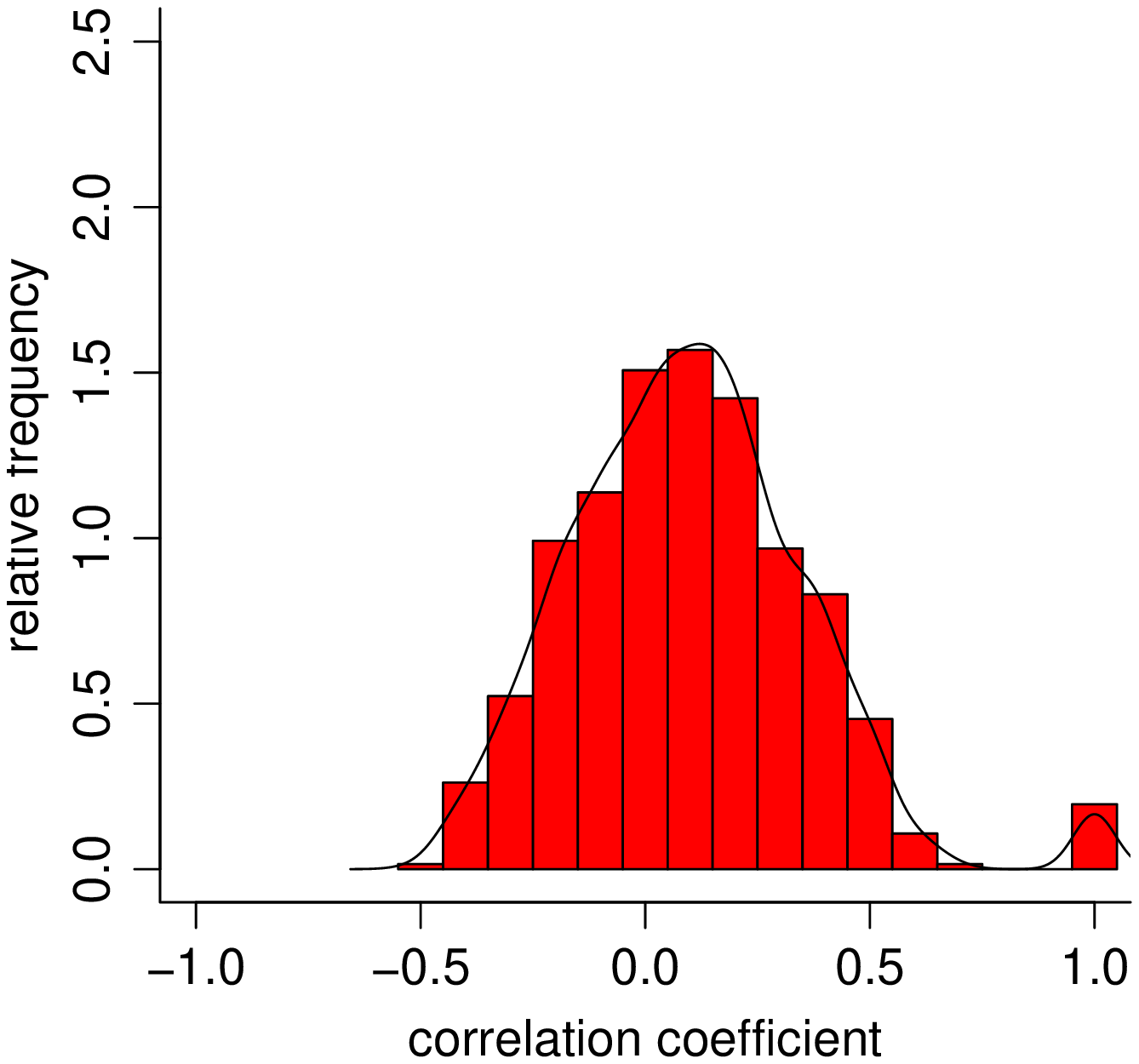}
}
\caption{Histogram of the correlation coefficients for 2MASS\,0847 from Fig.~\ref{2m0847_cor_rect_bin20_pair_relspec}.}
\label{2m0847_cor_bin20_histogram}
\end{figure}

The spectra Fig.~\ref{2m0847_spectra} and Fig.~\ref{zJ_spectra} show the \k1\ 
doublets, FeH features and water absorption bands, seen now in all four UCDs
observed. We can see the \k1\ line at 1.517\,\micron\ identified also in
SSSPM\,0828. From Figs.~\ref{2m0847_cor_rect_bin20_pair_relspec} and
\ref{2m0847_cor_bin20_histogram} we see no evidence for any correlated
variations in the relative spectra: the correlation matrix is hardly
distinguishable from the random one in
Fig.~\ref{ss0828_cor_bin20_pair_rnormspec}.  Interestingly, inspection of the
correlation matrices for the direct (non-relative) spectra does reveal
correlations within both the target and reference spectra.  However, this is
misleading, because we are observing through the Earth's variable atmosphere
so we expect to see variability in the direct spectra. We furthermore expect
this telluric variability to be correlated, e.g.\ different parts of the water
spectrum vary coherently as the water column density varies.
Little can be inferred from the direct spectra: A
simultaneous reference source must be used (and the analysis performed on the
relative spectra, as done throughout this paper).  Had we not not done this,
we would now be concluding that there are significant correlations in 2MASS\,0847.
This demonstrates that for variability studies it would be insufficient to
correct for a time-averaged telluric absorption using a standard star taken
just one or a few times per night.

\subsection{Correlations in spectral indices}

BJ02 defined some spectral indices intended to be both sensitive to dust cloud
variability in UCDs and minimally influenced by (telluric) water absorption.
Three of these bands have been formed with the present data (from the
rectified, paired, relative spectra -- see section~\ref{rectification}) by
integrating over the specified wavelength interval: j1 (1.05--1.09\,\micron);
j2 (1.17--1.30\,\micron); h1 (1.51--1.64\,\micron, the red end being truncated
with respect to the original definition); h2 (1.51--1.55\,\micron). For each
object, the correlations between these bands are measured. (Note that h1 and
h2 overlap so correlations with respect to these bands are not independent.)
The only object/bands which show a significant correlation (\modcor\,$>$\,0.8)
is SDSS\,0539 between j1 and h1 with $\rho_{j1,h1}=0.87$ (and $\rho_{j1,h2}=0.89$,
but this is not independent). From Figs.~2--4 of and Table 2 of BJ02
we see that j1 and h1 would be positively correlated under
dust-induced variability. On the other hand, so would j1 and j2, yet this is
not seen for SDSS\,0539 ($\rho_{j1,j2}=0.35$).

We may also form flux ratios, j1/j2, j1/h1 etc. which are ``unlogged'' colours, as
the rectification factor cancels out. That is, j1/j2 vs.\ j1/h1 (for example) examines the
correlation of j2 and h1 normalized relative to j1.
Of the 10 possible correlations between these per object (excluding h1/h2)
we see several significant correlations. For SSSPM\,0828 we have
$\rho_{j1/j2, j1/h1} = 0.92$ and
$\rho_{j1/h2, j1/j2} = 0.92$. 
The latter is the correlation plotted in Fig.~12 of BJ02 for a very similar
spectral type, L1.5V, where \cor\,=\,0.83. Moreover, Fig.~13 of the same paper shows negligible
correlation between j2/h2 and j1/j2, and for SSSPM\,0828 we have $\rho_{j2/h2, j1/j2}
= -0.09$. Thus these two objects show very similar variability patterns.
In SDSS\,0539 the one significant colour index is $\rho_{j1/j2, j2/h1} = -0.82$.
With 2MASS\,0559 we have $\rho_{j1/h1, j2/h1} = 0.84$ (0.96 when formed against h2
instead of h1).  

This analysis adds support to the conclusion of correlated variability, in
particular over a broad spectral range.

\subsection{Variability in the 1.31--1.49\,\micron\ water band}\label{water_var}

Nakajima et al.\ (2000) reported variability in a water band between 1.34 and
1.42\,\micron\ in a T dwarf (SDSS 1624+00) from a series of eight spectra
taken over a period of 80 minutes. As in the present work, they simultaneously
observed a reference star in the spectrograph slit.  They claim that the
variability is intrinsic to the T dwarf, although it is unclear if this
variability is {\em statistically} significant (such an analysis is not
reported) and the authors themselves acknowledge the dependence of this result
on a correct estimation of the photometric uncertainties.

From the \chisq\ plots in the present paper, this region appears to be
significantly variable for 2MASS\,0847, SSSPM\,0828 and SDSS\,0539 (but not for 2MASS\,0559).
However, the source counts are very low here and error sources not accounted
for in the basic noise model may dominate, rendering the \chisq\ statistic less
reliable.  The correlation analysis introduced in the present paper does not
assume a noise model for the data.  2MASS\,0847 shows no significant correlations
in this band, and SSSPM\,0828 only does at the blue edge. SDSS\,0539 and 2MASS\,0559 show
many correlated bins in this band, suggesting that there is intrinsic water
variability in at least these latter two targets.  There is also methane
band in the region 1.31--1.49\,\micron, so in principle the variability could
be due to this. This is plausible for 2MASS\,0559, which is a T5 dwarf according to
Burgasser et al.~\cite{burgasser00}, but perhaps even for SDSS\,0539 (an L5 dwarf
according to Fan et al.~\cite{fan00}) where we possibly see methane absorption
(section~\ref{SD0539}).

\section{Summary and Conclusions}\label{conclusions}

I have monitored four UCDs with differential infrared spectrophotometry to
look for evidence of correlated variability over timescales between 0.1 and
5.5\,hrs. 

Of the four targets monitored, three show significant evidence for positively
correlated variations. In the L2 dwarf SSSPM\,0828 the most significantly
correlated bands lie between 1010 and 1070\,nm, with the next most significant
regions (\modcor\,$>$\,0.8) coinciding with Fe, FeH, VO, \k1\ and water.
SSSPM\,0828
also shows similar broad band correlation patterns (and lack thereof) seen in
the L1.5V 2M1439 reported by BJ02 (observed from a different site with a
different instrument).  The other L2V monitored, 2MASS\,0847, is devoid of
any correlated variability.
The T5 dwarf 2MASS\,0559 shows well correlated (\modcor\,$>$\,0.8) variability at
925--962\,nm, 1347--1351\,nm and 1434--1448\,nm which is probably due to
water, possibly also methane. The L5 dwarf SDSS\,0539 is different from the other
three sources in that it exhibits many more regions of correlated
variability, larger correlation coefficients and some anticorrelated regions.
The latter coincide with TiO and FeH features.  Generally, however, the
correlations in SDSS\,0539 are quite broad band and cannot always be associated
with specific features. This would be consistent with variations in the dust
opacity.

I have demonstrated that the observed correlated variability is not due to OH
lines nor to telluric (water) absorption.  For example, much of the correlated
variability in 2MASS\,0559 and SDSS\,0539 occurs at wavelength regions where there is
little correlated variability in the direct (non-relative) reference star
spectra.

In conclusion, this work provides good evidence for correlated variability in
three of four ultracool dwarfs.  It is intrinsic to the sources and is not the
result of telluric variability or colour-dependent extinction effects
(Bailer-Jones \& Lamm \cite{bjl}) in the Earth's atmosphere (because the
relative fluxes are formed over very narrow bands).
The analysis does not require an estimate of the noise level (flux
amplitude), although the estimation of confidence intervals does use an
assumption on the shape of the noise distribution. This multiband correlation
analysis is more robust than single band analyses, because correlated
variations are more likely to be intrinsic.
This is not to say, however, that only wavelength-correlated variability in UCDs exists.
Some variability is seen in narrow UCD features, such as the \k1\ lines and
FeH, CrH, TiO, but it is not limited to readily-identified spectral features.
Much of the variability is broad band in nature, consistent with water and
perhaps even methane,
but also broader still, as would be expected by dust. Such a variety of
elements and chemical phases in the variability signature suggests a common
cause. A good candidate is local temperature and/or composition variations.
The opacity is very sensitive to
temperature, and local variations caused by rising convective columns of hot
gas (for example), would change the relative ratios of molecular and gas
species .

\section*{Acknowledgements}

The data analysis and plotting in this article has made extensive use of the
freely available R statistical package, http://www.r-project.org.  I am
grateful to its developers for the time and effort they have invested.  I
thank Viki Joergens, Bertrand Goldman and the anonymous referee for useful
discussions and comments.  The observations on which this work is based were
obtained during ESO programme 072.C-0275.


\end{document}